\documentclass[aps,superscriptaddress,twocolumn,nofootinbib]{revtex4}

\makeatletter

\usepackage{natbib}
\usepackage{dcolumn}
\usepackage{graphicx}
\usepackage{amsmath}
\usepackage[hang,small,bf,centerlast]{caption} 
\begin{document}
\title{A theory for long-memory in supply and demand}

\author{Fabrizio Lillo}
\affiliation{Santa Fe Institute, 1399 Hyde Park Road, Santa Fe, NM 87501}
\affiliation{INFM Unit\`a di Palermo and Dipartimento di Fisica e Tecnologie Relative, viale delle Scienze I-90128, Palermo, Italy}

\author{Szabolcs Mike}
\affiliation{Santa Fe Institute, 1399 Hyde Park Road, Santa Fe, NM 87501}
\affiliation{Budapest University of Technology and Economics,
H-1111 Budapest, Budafoki \'ut 8, Hungary}

\author{J. Doyne Farmer}
\affiliation{Santa Fe Institute, 1399 Hyde Park Road, Santa Fe, NM 87501}

\begin{abstract}
Recent empirical studies have demonstrated long-memory in the signs of
orders to buy or sell in financial markets \cite{Bouchaud04,Lillo03c}.  We show how
this can be caused by delays in market clearing.  Under the common
practice of order splitting, large orders are broken up into pieces
and executed incrementally.  If the size of such large orders is power
law distributed, this gives rise to power law decaying autocorrelations in the
signs of executed orders.  More specifically, we show that if the
cumulative distribution of large orders of volume $v$ is proportional
to $v^{-\alpha}$ and the size of executed orders is constant, the autocorrelation 
of order signs as a function of the lag $\tau$ is asymptotically proportional to $\tau^{-(\alpha - 1)}$.  This is a long-memory process when $\alpha < 2$.  With a few caveats, this gives
a good match to the data.  A version of the model also shows long-memory fluctuations in order execution rates,  which may be relevant for explaining the long-memory of price diffusion rates.
\end{abstract}
\maketitle 
\tableofcontents
\section{Introduction}

A random process is said to have long-memory if it has an autocorrelation function that is not integrable. This happens, for example, when the autocorrelation function decays asymptotically as a power law of the form $\tau^{-\gamma}$ with $\gamma<1$. This is important because it implies that values from the distant past can have a significant effect on the present, that the stochastic process lacks a typical time scale, and implies anomalous diffusion in a stochastic process whose increments have long-memory. Examples of long-memory processes and anomalous diffusion have been observed in many physical, biological and economic systems ranging from turbulence \cite{Richardson26} to chaotic dynamics due to flights and trapping \cite{Geisel85}, dynamics of aggregates of amphiphilic molecules \cite{Ott90} and DNA sequences \cite{Peng92,Peng94}. In finance the volatility, roughly defined as the diffusion rate of price fluctuations, is known to be a long-memory process \cite{Ding93,Breidt93}.  In this paper we analyze a mechanism for creating a long-memory process, based on converting a static power law distribution into a random process with a power law autocorrelation function.  Other examples of stochastic processes relating power laws to long-memory have been given by Mandelbrot \cite{Mandelbrot69} (analyzed by Taqqu and Levy \cite{Taqqu86}),  and in the context of DNA sequences by Buldyrev et al. \cite{Buldyrev93}.  

Recently a new long-memory property of the order flow in a financial market was independently observed by Bouchaud et al. in the Paris Stock Exchange \cite{Bouchaud04} and Lillo and Farmer in the London Stock Exchange (LSE) \cite{Lillo03c}.  These studies have shown that there is a remarkable persistence in buying vs. selling.  Labeling the signs of trading orders as $\pm 1$ according to whether they are to buy or to sell, the autocorrelation of observed order signs is strongly positive,  asymptotically decaying roughly as a power law $\tau^{-\gamma}$, where $\gamma \approx 0.6$.  Such positive autocorrelations can be measured at statistically significant levels over time lags as long as two weeks.  

\begin{figure}[ptb]
\includegraphics[scale=0.3,angle=-90]{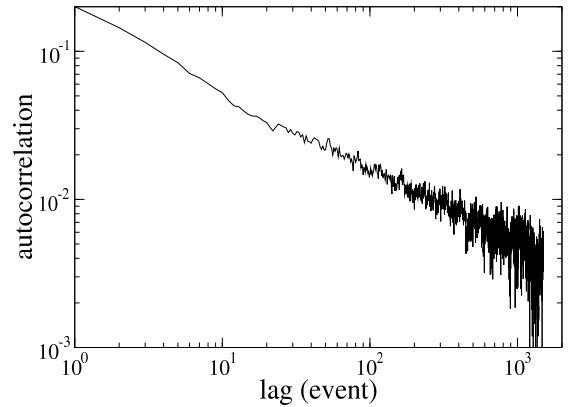}
\caption{Autocorrelation function of the time series of signs of orders that result in immediate trades for the stock Shell traded at the London Stock Exchange in the period May 2000 - December 2002, a total of $5.8 \times 10^5$ events.}
\label{shell}
\end{figure}

For example, in Fig.~\ref{shell} we show the empirical autocorrelation function of the time series of signs of orders that result in immediate trades for the stock Shell. The autocorrelation function is well described by a power law decay over almost three decades and a least squares fit to this gives $\gamma=0.53$. The fact that $\gamma < 1$ implies that this is a long-memory process, i.e. its autocorrelation function decays so slowly that it is not integrable.  This is important because it implies that values from the distant past have a significant effect on the present.  A diffusion process built from long-memory increments has a variance $\sigma^2$ that grows in time as $\sigma^2(\tau) \sim \tau^{2H}$, where is called the Hurst exponent.  For $0 < \gamma < 1$, $H = 1 - \gamma/2$.  For a normal diffusion process $H = 1/2$, but when $H > 1/2$ the variance grows faster than $\tau^{1/2}$, which is called anomalous diffusion.  Another important consequence is that statistical averages converge slowly, e.g. the mean of a quantity that displays anomalous diffusion converges as $T^{-(1-H)}$, where $T$ is the sample size.  The signs of orders in the LSE have been shown to pass tests for long-memory with a high degree of statistical significance \cite{Lillo03c}. 

From an economic point of view this is important because of its implications for market efficiency.  All other things being equal, since buy orders tend to drive the price up and sell orders tend to drive them down, this would imply that it was possible to make profits using a simple linear model to predict future price moments.  In order to prevent this the market has to make substantial compensating adjustments \cite{Bouchaud04,Lillo03c,Bouchaud04b}.  The difficulty of making such adjustments perfectly may have important implications about the origin of long-memory in the volatility of prices.

In this paper we hypothesize that the cause of the long-memory of order flow is a delay in market clearing. To make this clearer, imagine that a large investor like Warren Buffet decides to buy ten million
shares of a company. It is unrealistic for him to simply state his demand to the world and let the market do its job.  There are unlikely to be sufficient sellers present, and even if there were, revealing a large order tends to push the price up.  Instead he keeps his intentions as secret as possible and trades the order incrementally over an extended period of time, possibly through intermediaries.  In a study of this phenomenon, about a third of the dollar value of such institutional trades took more than a week to complete \cite{Chan93,Chan95}.  This conflicts with standard neoclassical economic models, which assume market clearing, i.e. that the price always adjusts so that supply and demand are evenly matched.   The fact that large orders are kept secret and executed incrementally implies that at any given time there may be a substantial imbalance of buyers and sellers, which can be interpreted as a failure of market clearing.   Supply and demand do not match, and the market fails to clear.  Effective market clearing is delayed, by variable amounts that depend on fluctuations in the size and signs of unrevealed orders.

We propose a simple model to explain the long-memory of order flow based on
delays in market clearing. We postulate that unrevealed {\it hidden orders}
are distributed according to a power law.  These are broken up into
pieces, which we call {\it revealed orders}, that are submitted at a steady rate.
We show that this leads to long-memory in order flow, yielding a model consistent with empirical observations.  The main result is an analytic computation relating the exponent of the power law of
the volume distribution of hidden orders to the rate of decay of the long-memory
process characterizing revealed orders.

The paper is organized as follows:  In Section~II we define the two models that we study here, which we call the fixed $N$ model and the $\lambda$ model.  In Section~III we analytically compute the autocorrelation function of revealed orders for the fixed $N$ model in terms of the parameters, and test it against simulation results.  Section~IV discusses the properties of the $\lambda$ model, showing that it displays interesting temporal fluctuations.  Section~$V$ compares the predictions to empirical evidence and discusses the assumptions of the model in the context of real markets.  In Section~VI we discuss the possible broader implications.

\section{Description of models}

We develop a model with two variations, which we call the $\lambda$~{\it model} and the {\it fixed N~model}.  We first describe the $\lambda$ model, which is more realistic, but for which we have
only simulation results.  We then describe the fixed $N$ model, which
is less realistic, but has the important advantage of being simpler,
allowing us to obtain analytic results.  Because of the simple nature of these results, they apply equally well to the $\lambda$~model.

We first describe the $\lambda$~model.  Let $N(t)$ be the number of
hidden orders at time $t = 1,2, \ldots, T$.  At each time $t$ generate
a new hidden order with probability $0 < \lambda < 1$ if $N(t) > 0$, or
probability one if $N(t) = 0$.  Assign each new hidden order a random
sign $s_i$ and an initial size $v_i(t^*) = L \Delta v$, where $t^*$ is the time when the hidden order is created, and $L = 1,2, \ldots$ is drawn from a Pareto distribution $P(L) = \alpha L^{-(\alpha + 1)}$, with $\alpha > 0$.  The random variables $L$ and $s_i$ are IID\footnote{In the language of extreme value theory \cite{Embrechts97}, the Pareto distribution is just one example of a power law.  A distribution $f(x)$ is a power law with tail exponent $\alpha$ if there exists a slowly varying function $g(x)$ such that $\lim_{x \to \infty} f(x) g(x) = K x^{-\alpha}$, where $K$ and $\alpha$ are
positive constants.  A function $g(x)$ is a slowly varying function if for any $t > 0$ $\lim_{x \to \infty} g(tx)/g(x) = 1$.  A common example of a slowly varying function is $\log x$,  so in this sense the function $x^{-\alpha}\log x$ is a power law.  Thus, the term ``power law'' refers not to a specific distribution, but to an equivalence class of distributions with the same asymptotic scaling properties.  It is clear from the calculations leading up to our main result, equation~(\ref{paretoScaling}), that it is not necessary to assume that the distribution of volumes is strictly Pareto distributed; any power law distribution $p(L)$ with a given tail exponent $\alpha$ will give the same asymptotic scaling for the autocorrelation function of revealed orders.}.  At each timestep $t$ an existing hidden order $i$ is
chosen at random with uniform probability, and a volume $\Delta v$ of
that order is removed, so that $v_i(t+1) = v_i(t) - \Delta v$.  This
generates a revealed order of volume $\Delta v$ and sign $x_t = s_i$.  A
hidden order $i$ is removed if $v_i(t+1) = 0$.  Thus, the number of
hidden orders $N(t)$ fluctuates in time, depending on fluctuations in
arrival and removal. 

The fixed $N$ model is the same, except that the number of hidden
orders $N$ is kept fixed.  Thus, if a hidden order is removed it is
immediately replaced by a new one with a random sign and a new size. 

The main result of this paper is the calculation of the
autocorrelation function of revealed order signs $x_t$ for the fixed $N$
model.  We show in the next section that the tail of the
autocorrelation function asymptotically scales as $\tau^{-(\alpha - 1)}$.  While
varying $N$ affects the shape of the autocorrelation function for
small $\tau$, providing $\alpha$ is held fixed, it does not affect its
asymptotic scaling. Even though $N(t)$ varies in the $\lambda$ model, the asymptotic behavior is independent of $N(t)$, and so the asymptotic behavior of the autocorrelation function is the same. This
is particularly convenient because it allows us to make a prediction
in terms of observable quantities (see Section~\ref{testing}).

\section{Analytic computation for fixed~$N$ model}

Because the hidden order arrival process is IID, it is possible to
compute the autocorrelation of the fixed $N$ model analytically.  The
basic idea of the computation is to understand the behavior of the
autocorrelation conditioned on $L$, the initial length of the hidden
order in units of the revealed order size $\Delta v$, and then combine
the results for different values of $L$. 

We first begin by giving a simple intuitive argument for the asymptotic scaling. The probability at any instant of time that a revealed order comes from a hidden order of length $L$ is $Q(L) \propto Lp(L)$. This revealed order contributes to inducing a positive autocorrelation at lag $\tau$ only if the revealed order $\tau$ steps ahead comes from the same hidden order. In other words, in order to contribute to the autocorrelation function at lag $\tau$, a hidden order must be of length $L > A \tau$, where $A$ is a constant.  Summing over all hidden orders gives an autocorrelation  $\rho(\tau) \sim \int_{A \tau}^\infty Q(L) \sim \tau^{-(\alpha-1)}$, which is the main result of Eq.~\ref{paretoScaling}.  In the remainder of this section, we present a more detailed calculation, which also allows us to compute the correct prefactor.

\subsection{Autocorrelation in probabilistic terms}

Under the convention that
the signs of the revealed orders are $x_t = \pm 1$, because of the
symmetry between buying and selling $E[x_t] = 0$ and $E[x_t^2] =
1$, where $E$ denotes the expectation. Therefore the autocorrelation is simply $\rho(\tau)=E[x_t
x_{t+\tau}]$.  We can rewrite this as
\begin{equation}
E[x_t x_{t+\tau}]=\sum_{L = 1}^{\infty}\ Q(L) E[x_t x_{t+\tau}| L],
\end{equation}
where $E[x_t x_{t + \tau} | L]$ is conditioned on the hidden order
that generated $x_t$ having length $L$.  $Q(L)$ is the probability
that a revealed order drawn at random comes from a hidden order of
length $L$.  Let $q(\tau|L)$ be the probability that revealed orders
at times $t$ and time $t + \tau$ came from the same hidden order,
given that it has original length $L$.  Because $E[x_t x_{t + \tau}] =
0$ if $x_t$ and $x_{t+\tau}$ came from different hidden orders, and
$E[x_t x_{t + \tau}] = 1$ if they came from the same hidden order, the
conditional expectation can be rewritten
\begin{equation}
E[x_t x_{t+\tau}|L] = q(\tau|L),
\end{equation}
which implies
\begin{equation}
\rho(\tau)=\sum_{L=1}^\infty Q(L) q(\tau|L). 
\end{equation}

To compute $Q$, we note that the number of revealed orders coming from
hidden orders of length $L$ is proportional to $L p(L)$, where
$p(L)$ is the probability that a hidden order has length $L$.  To
compute $Q(L)$ we must properly normalize this by summing over
$L$,
\begin{equation}
Q(L) = \frac{L p(L)}{\sum_{L=1}^\infty L p(L)}.
\label{Q}
\end{equation}
 This gives
\begin{equation}
\rho(\tau)=\frac{1}{\bar L}\sum_{L=1}^{\infty}  L q(\tau|L) p(L),
\label{rho}
\end{equation}
where $\bar L$ is the average value of $L$.

The conditional probability $q(\tau|L)$ can be written
\begin{equation}
w(L, \tau) p,
\end{equation}
where $w (L, \tau)$ is the probability that a given hidden order
is still active after time $\tau$, and $p$ is the probability that
it will be selected for execution assuming it is still active.  By
assumption $p = 1/N$.  

Computing $w(L, \tau)$ is more complicated: Let $s$ be the number of
revealed orders drawn from a given hidden order during the $\tau - 1$
timesteps between time $t$ and time $t + \tau$, and let $P_{\tau -
  1}(s < k)$ be the probability that $s$ is less than a given value
$k$.  Thus, for a hidden order that has length $l$ at time $t$, the
probability that it still exists at time $t + \tau$ is $P_{\tau - 1}
(s < l)$.  For a hidden order with original length $L$, $l$ is
uniformly distributed with probability $1/L$ over the values $1,
\ldots, L$.  Thus we can express $w(L, \tau)$ as a sum of
probabilities, one for each possible value of $l$.
\begin{eqnarray}
w(L, \tau) =\frac{1}{L}(P_{\tau-1}(s< L-1)+P_{\tau-1}(s<
L-2)+\nonumber\\ \ldots + P_{\tau-1}(s< 1)).
\end{eqnarray}

The probabilities $P_{\tau-1}(s< k)$ can be expressed as sums of
binomial probabilities, corresponding to the possible sequences with
which a given hidden order generates $k-1$ revealed orders.
\begin{equation}
P_{\tau-1}(s< k)=\sum_{h=0}^{k-1}{\tau-1 \choose h} p^h (1-p)^{\tau-1-h}.
\end{equation}
Therefore
\begin{equation}
q(\tau|L)=\frac{p}{L}\sum_{j=1}^{L-2}\sum_{h=0}^j{\tau-1
  \choose h}p^h(1-p)^{\tau-1-h}.
\label{q}
\end{equation}

\subsection{de Moivre-Laplace approximation}

The autocorrelation can now be computed using Eq.~(\ref{rho}).
However, since the sums of binomial coefficients are difficult to
manage we will make use of the de Moivre-Laplace approximation
\cite{Feller50}. For $npq>>1$ one can approximate
\begin{equation}
{n \choose k} p^k q^{n-k}\simeq \frac{1}{\sqrt{2\pi npq}}
\exp\left(-\frac{(k-np)^2}{2npq}\right).
\end{equation}
As a consequence the sum of consecutive terms of a binomial
distribution can be approximated as
\begin{eqnarray}
\sum_{k=k_1}^{k_2}{n \choose k} p^k q^{n-k}\simeq \\\frac{1}{2}\left[{
    {\rm erf}}\left(\frac{k_2-np+1/2}{\sqrt{2npq}}\right)-{{\rm erf}}\left(\frac{k_1-np-1/2}{\sqrt{2npq}}\right)\right],\nonumber
\end{eqnarray}
where ${\rm erf}$ is the error function.

By converting the sum to an integral, and letting $s = \tau - 1$,
equation~(\ref{q}) becomes
\begin{widetext}
\begin{eqnarray}
q(s + 1|L)\simeq \frac{p}{2L}\sum_{j=1}^{L-2}\left[{
    {\rm erf}}\left(\frac{j-sp+1/2}{\sqrt{2sp(1-p)}}\right)-{{\rm erf}}\left(\frac{-sp-1/2}{\sqrt{2sp(1-p)}}\right)\right]\simeq\nonumber\\
\frac{p}{2L}\int_{1/2}^{L-2+1/2}\left[{
    {\rm erf}}\left(\frac{x-sp+1/2}{\sqrt{2sp(1-p)}}\right)-{{\rm erf}}\left(\frac{-sp-1/2}{\sqrt{2sp(1-p)}}\right)\right]~dx,
\end{eqnarray}
\end{widetext}
For the approximation of the sum by the integral we use
$\sum_{i=a}^bf(i)\simeq\int_{a+1/2}^{b+1/2}f(x)dx$. Performing the
last integral gives
\begin{widetext}
\begin{eqnarray}
q(s + 1|L)\simeq
\frac{p}{2L}\huge(-\exp(-\frac{(sp)^2}{2sp(1-p)})+\sqrt{\frac{2}{\pi}}\sqrt{sp(1-p)}(\exp(-\frac{(L-1-sp)^2}{2sp(1-p)}))+
\nonumber\\(sp-1){\rm erf}(\frac{1-sp}{\sqrt{2sp(1-p)}})+(L-2){\rm erf}(\frac{1/2+sp}{\sqrt{2sp(1-p)}})+(1+sp-L){\rm erf}(\frac{1-L+sp}{\sqrt{2sp(1-p)}})\big).
\label{qfinal}
\end{eqnarray}
\end{widetext}
The sum over $L$ in Eq.~(\ref{rho}) can be approximated by the integral
\begin{equation}
\rho(\tau)\simeq\int_{1+1/2}^{\infty}q(\tau|L)\frac{p(L)L}{\bar L}~dL.
\label{final}
\end{equation}

Finally, we need to translate the domain of validity of the de
Moivre-Laplace approximation into more relevant terms.  The condition
$npq>>1$ in Eq.~(\ref{q}) becomes $(\tau-1)p(1-p)>>1$.  This
leads to the condition
\begin{equation}
\tau>>\frac{N^2}{N-1}-1\simeq N, 
\label{validity}
\end{equation}
i.e. the approximation is valid as long as the lag is much greater
than the number of hidden orders.  Since the number of hidden orders
is fixed, the approximation is always valid for sufficiently large
$\tau$.

We have tested these calculations for the simple case in which all
hidden orders have the same size $L_0$, i.e.  $p(L)=\delta(L-L_0)$,
where $\delta$ is the Dirac delta function.  This implies $\rho(\tau)
= q(\tau | L_0)$, so that Eq.~(\ref{qfinal}) gives a closed form
expression for the autocorrelation function.  As expected, the approximation
always agrees very well for large values of $\tau$.  The agreement is
also good for small values of $\tau$ when $N$ is small and $L_0$ is
sufficiently large.

\subsection{Pareto distribution}

We now consider the more realistic case that the hidden order size $L$
has a Pareto distribution
\begin{equation}
p(L)=\frac{\alpha}{L^{\alpha+1}}
\label{pareto}
\end{equation}
where $\alpha >1$ is the tail exponent. In this case the integral of
Eq.~(\ref{final}) cannot be performed analytically.  We can, however,
give an analytical asymptotic expansion of the integral
(\ref{final}). The calculations detailed in the appendix
make use of the saddle point approximation. The result is
that the leading term of the asymptotic expansion of $\rho(\tau)$ is
given by the terms depending on ${\rm erf}$ functions in Eq.~(\ref{qfinal}),
and the autocorrelation function decays asymptotically as
\begin{equation}
\rho(\tau) \sim \frac{N^{\alpha-2}}{\alpha}\tau^{-(\alpha - 1)}.
\label{paretoScaling}
\end{equation}
This result indicates that the autocorrelation function decays as a
power law with exponent $\gamma = \alpha - 1$.  The number of
hidden orders affects the prefactor, but does not affect the scaling
exponent.  Interestingly, when $\alpha = 2$ the prefactor is independent of $N$.  When $\alpha < 2$ it is a decreasing function of $N$, and when $\alpha > 2$ it is an increasing function of $N$. The value $\alpha=2$ separates the regime where the size of hidden orders has infinite variance from the regime where the variance is finite\footnote{Note that Buldyrev et al. \cite{Buldyrev93} found a similar formula in the context of structure in DNA sequences.}.

\begin{figure}[ptb]
\begin{center}
\includegraphics[scale=0.35]{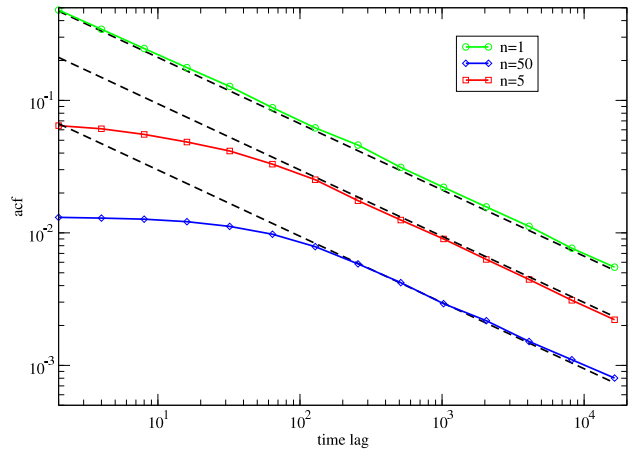}
\end{center}
\caption{(Color online) Autocorrelation of the fixed $N$ model with $\alpha = 1.5$,
  for $N = 1$ (green circles), $N = 5$ (red squares) and $N = 50$ (blue diamonds), based on a
  simulation with $T = 10^9$.  This is compared to the asymptotic predictions of Eq.~(\ref{paretoScaling}), shown as dashed black lines.}
\label{comparison}
\end{figure}

Fig.~\ref{comparison} compares the autocorrelation function predicted by Eq.~(\ref{paretoScaling}) to a simulation for $\alpha = 1.5$, $N = 1$, $N = 5$, and $N = 50$.  
For large values of $\tau$ the match is excellent, both in terms of the slope and the size of the prefactor.  For $N = 1$ the prediction matches the simulation across the entire range of $\tau$.  As expected, when $N$ increases the prediction deviates at small $\tau$, but still matches for large $\tau$.  We have also checked the consequences of varying $\alpha$ and find that the prefactor behaves as predicted by Eq.~(\ref{paretoScaling}).

Note that we used $T=10^9$ samples to simulate the model and compare to theory.  This is because for $\alpha = 1.5$ this is a strongly long-memory process, and the convergence is extremely slow.  This will become an issue later on when we test the model against real data -- even for very large sample sizes the error bars remain quite large.

\section{Liquidity fluctuations of the $\lambda$~model}

\begin{figure}[ptb]
\begin{center}
\includegraphics[scale=0.5]{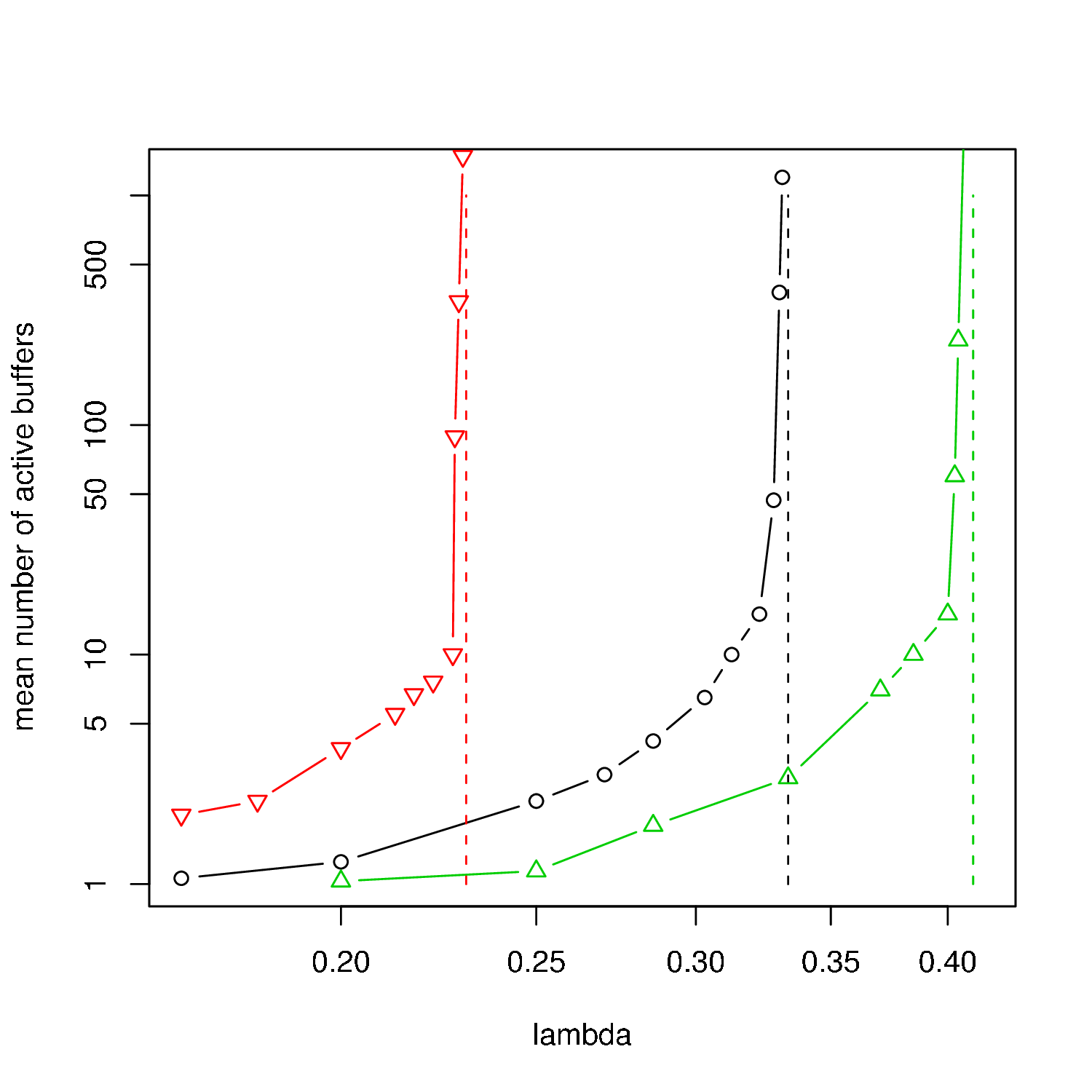}
\end{center}
\caption{(Color online)The average number of hidden orders as a function of the creation parameter $\lambda$ for
$\alpha = 1.3$ (red downward pointing triangles), $\alpha = 1.5$ (black circles) and $\alpha = 1.7$ (green upward pointing triangles).  The dashed lines are the corresponding predicted critical values $\lambda_c = (\alpha - 1)/\alpha$.}
\label{critical}
\end{figure}

We now return to discuss the $\lambda$ model.  As a reminder, this differs from the fixed $N$ model analyzed so far in that the number of buffers $N(t)$ is not fixed.  Instead, new buffers are added with probability $\lambda$ when $N(t) > 0$, and probability $1$ otherwise.  For the mean of $N(t)$ to remain bounded it is necessary that the rate of creation of new orders equal the rate at which they are removed.  This implies the model has a critical threshold where $E[N(t)] \to \infty$.  This can be simply computed as follows:  Let $n(t)$ be the total number of future revealed orders stored in all hidden orders at time $t$, i.e. $n(t) = \sum_{i=1}^{N(t)} v_i(t) /\Delta v $.  The average rate of change of $n(t)$ is
\begin{equation}
E[n(t+1) - n(t)] = R(n(t)) \bar{L} - 1.
\nonumber
\end{equation}
The first term represents addition of a new hidden order, and the second term the removal of a revealed order at every timestep.  The creation rate $R(n(t)) = \lambda$ when $n(t) > 0$ and $R(n(t)) = 1$ otherwise.  The average length of a new hidden order is $\bar{L}$, which under the Pareto assumption is $\bar{L} = \sum_{L=1}^\infty L (L) = \alpha/(1 - \alpha)$.  In the limit where $E[n(t)]$ is large it is a good approximation to say that $n(t)$ is never zero, so that $R(n(t)) = \lambda$.  Setting $E[n(t+1) - n(t)]  = 0$ implies the critical value $\lambda_c$ is
\begin{equation}
\lambda_c = 1/\bar{L} = (\alpha - 1)/\alpha = \gamma/\alpha.
\end{equation}
For the last equality we have made use of the fact that $\gamma$ does not depend on $N$ in Eq.~(\ref{paretoScaling}), which indicates that $\gamma = \alpha - 1$ applies equally well to the $\lambda$ model as long as $\lambda < \lambda_c$ (we have verified this in simulations).  We also confirm the dependence of the critical behavior on $\alpha$ in Fig.~\ref{critical}.

\begin{figure}[ptb]
\begin{center}
\includegraphics[scale=0.3]{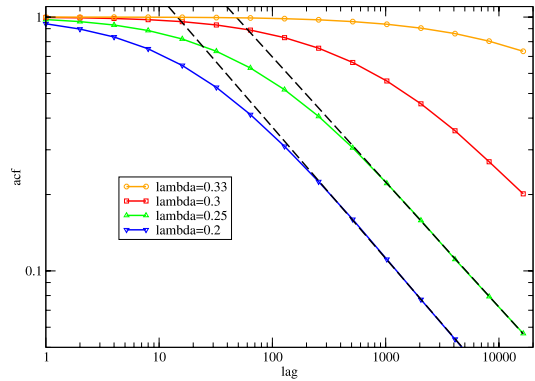}
\end{center}
\caption{(Color online) Autocorrelation function of the number of active hidden orders in the $\lambda$~model for four different values of $\lambda$, as shown in the inset.  The dashed black lines have slope $\alpha - 1$.}
\label{Nautocor}
\end{figure}

One of the interesting features of the $\lambda$~model is that it generates long-memory fluctuations in the number of active hidden orders.   This is caused by positive feedback between the number of orders and the accumulation rate.  This is because the average rate at which hidden orders are executed is $1/N(t)$.  Thus when $N(t)$ is larger than average, the rate at which active hidden orders are removed is lower than average, which tends to cause $N(t)$ to increase above its average value.  Such an increase is triggered by random fluctuations in which one or more particularly large orders are created; when these orders are finally removed, $N(t)$ decreases.  $N(t)$ thus makes large and persistent fluctuations.  The autocorrelation function has an asymptotic power law decay of the form $\rho_N(\tau) \sim \tau^{-\gamma}$ as shown in Fig.~\ref{Nautocor}.  From simulations, we find that $\gamma = \alpha - 1$.

For this model fluctuations in the number of hidden orders correspond to fluctuations in the time to execute an order.  In economics this is one aspect of what is called {\it liquidity}, which is a general term referring to the ease of execution of an order. One of the interesting properties of prices of economic time series is that they display what is commonly called {\it clustered volatility}, i.e. the diffusion rate of price changes is strongly autocorrelated in time, and in fact is a long-memory process \cite{Ding93,Breidt93}.  It has recently been shown that this is related to fluctuations in liquidity, in this case defined as the price response to an order of a given size \cite{Farmer04b}.  The fact that this kind of model predicts long-memory fluctuations in another aspect of liquidity (the time to execute an order) may be related to the explanation of clustered volatility.

\section{Testing the predictions\label{testing}}

Unfortunately, data comparing hidden orders and revealed orders are not widely available, which complicates the problem of testing this model.  The only data set we know of that includes the kind of data that is needed for a proper test was used by Chan and Lakonishok \cite{Chan93,Chan95} to study the execution of customer orders at large brokerage firms.  Unfortunately, they did not fit functional forms to the size distributions or test for long-memory, and we have not been able to obtain their data.  Their study does make it clear that order splitting is very common, and suggests that the time scale on which order splitting occurs is sufficiently long to match the autocorrelations in order flow.

We compare the predictions of the model to the data in two different ways.  The first is based on computation of the scaling exponents, described in Section~\ref{gamma}, and the second is based on the properties of run length, described in Section~\ref{runlength}.  Before presenting the first test, we must first review the market structure.

\subsection{Market structure and order distributions}

Although we have no transaction data with direct information about hidden orders, we can perform an indirect test of the scaling relations predicted by the model which takes advantage of the market structure used in the New York Stock Exchange and the London Stock Exchange.  They both employ two parallel markets which provide alternative methods of trading, called the on-book or ``downstairs'' market, and the off-book or ``upstairs'' market.  In the LSE orders in the on-book market are placed publicly but anonymously and execution is completely automated.  The off-book market, in contrast, operates through a bilateral exchange mechanism, via telephone calls or direct contact of the trading parties.  The anonymous nature of the on-book market facilitates order splitting, and it is clear that it is a common practice.  This is also supported by the fact that in our data set it is possible to track the on-book orders for individual trading institutions, and the long-memory property of order flow is evident even for single institutions \cite{Lillo03c}.  In contrast, off-book trading is based on personal relationships and order splitting is believed to be less frequent.  This is because a series of orders of the same sign tend to gradually change the price in a direction that is unfavorable to the other party \cite{Chan93,Chan95}.

Thus one might make the hypothesis that in the off-book market people just submit their orders rather than hiding them, while in the on-book market they hide their true orders and execute them through a series of revealed orders.  While there is some truth in this hypothesis, it is not strictly true.  When we examine sequences of off-book trades for individual institutions, we often see long runs of trades of the same sign, suggesting that order splitting is also fairly common in the off-book market.  Even though order-splitting is not common when trading with the same party, it is still possible to split a large order and trade it in the off-book market with many different parties.  Thus the transactions in the off-book market have already undergone some order splitting, and it is not clear how well the distribution of transactions corresponds to that for hidden orders.  

Despite the caveats mentioned above, we will press forward with the hypothesis that off-book trades can be used as a proxy for hidden orders, and see how the predictions of our model match the empirical observations of order splitting.  To this end we select $20$ highly capitalized stocks traded at the London Stock Exchange in the period May 2000 - December 2002.  The stocks we analyzed are Astrazeneca (AZN), Bae Systems (BA.), Baa (BAA), BHP Billiton (BLT), Boots Group
(BOOT), British Sky Broadcasting Group (BSY), Diageo (DGE), Gus (GUS),
Hilton Group (HG.), Lloyds Tsb Group (LLOY), Prudential (PRU), Pearson
(PSON), Rio Tinto (RIO), Rentokil Initial (RTO), Reuters Group (RTR),
Sainsbury (SBRY), Shell Transport \& Trading Co. (SHEL), Tesco (TSCO),
Vodafone Group (VOD), and WPP Group (WPP).  The number of trades for the combined group of stocks is $16.7 \times 10^6$; of these $11 \times10^6$ are on-book trades and $5.7 \times10^6$ are off-book trades.

\begin{figure}[ptb]
\begin{center}
\includegraphics[scale=0.3,angle=-90]{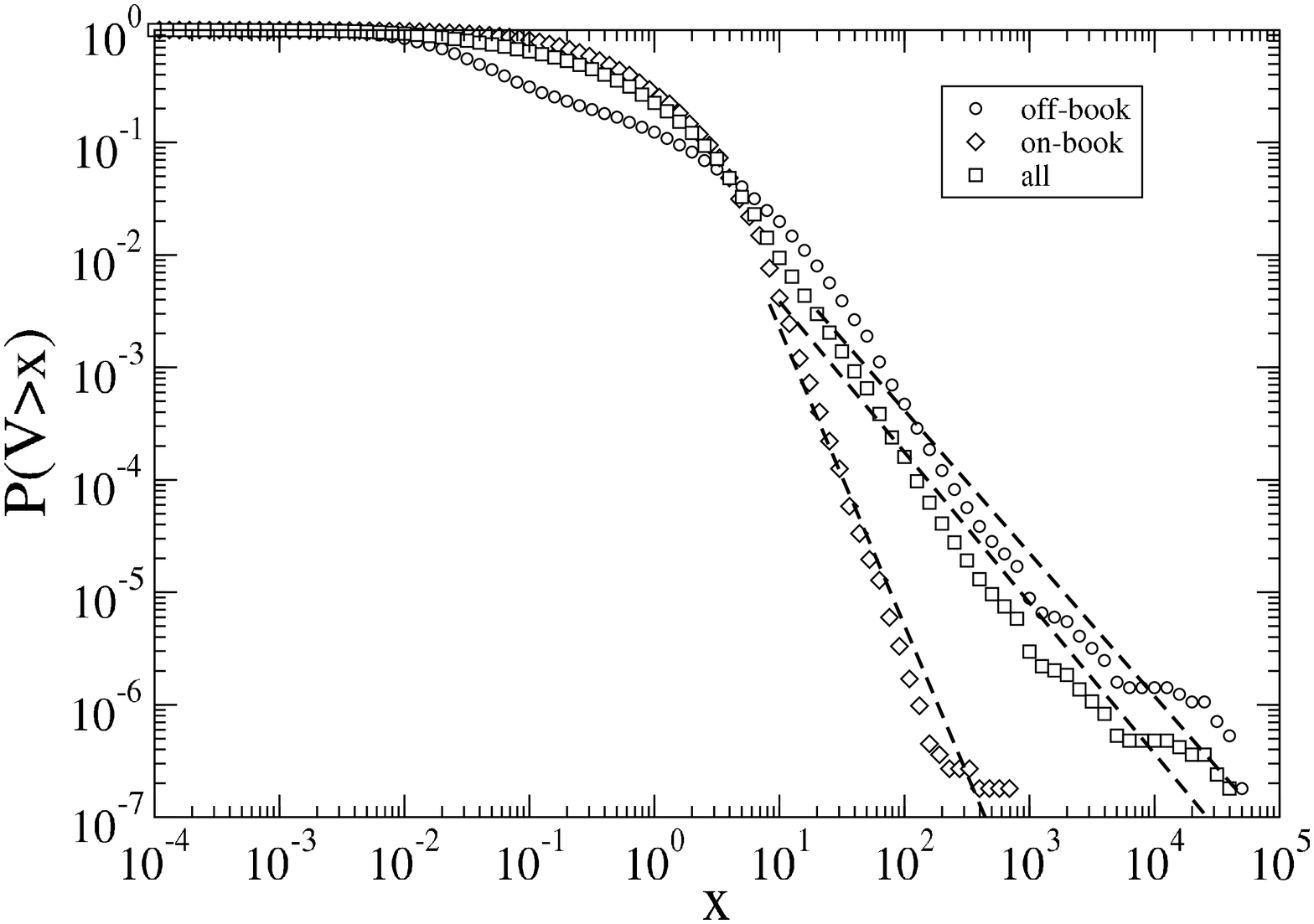}
\vskip 0.3in
\includegraphics[scale=0.3,angle=-90]{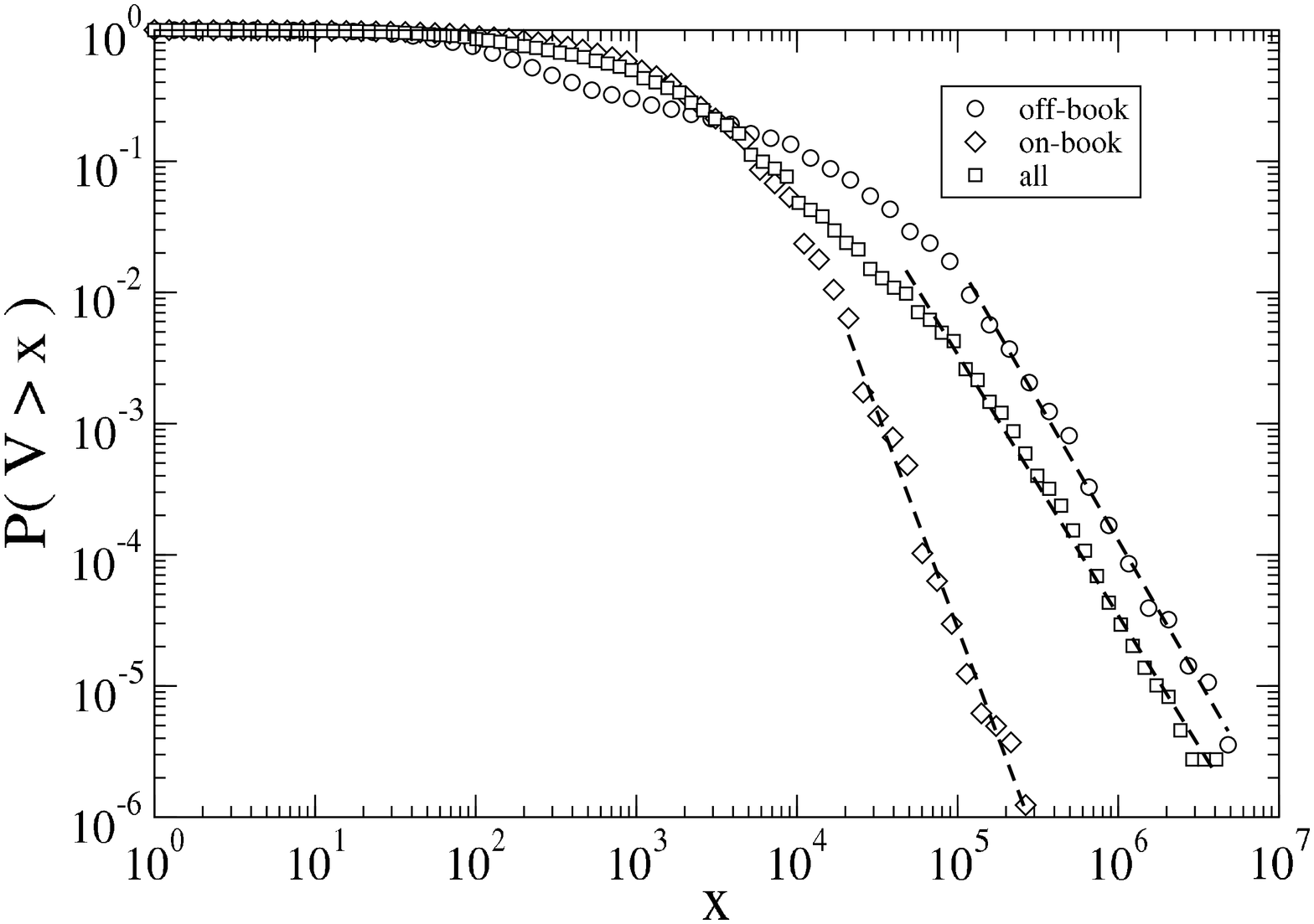}
\caption{Volume distributions of off-book trades (circles), on- book trades (diamonds) and the aggregate of both (squares).  In (a) we show this for a collection of 20 different stocks, normalizing the volume of each by the mean volume before combining, whereas (b) shows unnormalized values (in shares) for the stock Astrazeneca.  The number of trades in each case is $11 \times 10^6$ (aggregate on-book), $5.7 \times 10^6$ aggregate off-book, $8.0 \times10^5$ (AZN on-book) and $2.8 \times10^5$ (AZN off-book).  The dashed black lines have the slope found by the Hill estimator (and are shown for the largest one percent of the data).}
\label{volumeDist}
\end{center}
\end{figure}

\begin{figure}[ptb]
\begin{center}
\includegraphics[scale=0.3,angle=-90]{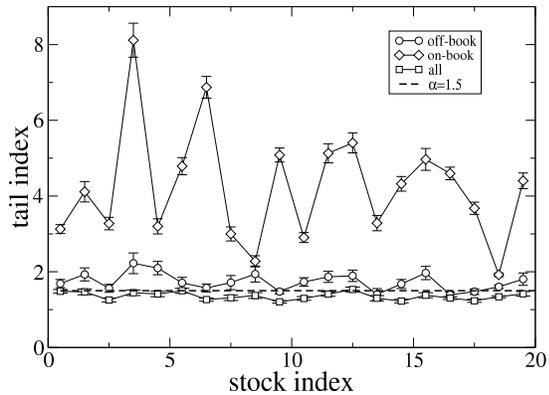}
\caption{Scaling exponents $\alpha$ for the twenty stocks we study here, based on the hypothesis that the largest one percent of the trades $V$ are described by the relation $P(V > x) \sim x^{-\alpha}$.  The stocks are arranged along the $x$ axis in alphabetical order.  The circles refer to off-book trades, the diamonds to on-book and the squares to the aggregate of both. For comparison we draw a dashed line for $\alpha = 1.5$.}
\label{tailExponents}
\end{center}
\end{figure}

In Fig.~\ref{volumeDist} we show the empirical probability distributions for the volume of trades in both the off-book and on-book markets in the London Stock Exchange.  We show this for an aggregate of 20 heavily traded stocks and for the single stock Astrazeneca, which is typical of the stocks in the sample.  This makes it clear that the tails die out more slowly in the off-book market.  The largest trade sizes in the off-book market are more than a factor of ten larger than those in the on-book market; for Astrazeneca, for example, the largest orders are roughly four million shares in the off-book market vs. 200 thousand in the on-book market.  Alternatively, to measure the decay of the tails more quantitatively, we assume the asymptotic relation for volume $V$ is $P(V > x) \sim x^{-\alpha}$, and estimate $\alpha$ using a Hill estimator applied to the largest one percent of the data \cite{Hill75}.  For the aggregate data set this gives $\alpha = 1.59$ for the off-book data, $\alpha = 2.90$ for the on-book data, and $\alpha = 1.64$ for the combined data\footnote{The results for the combined data set are in rough agreement with those first reported for the NYSE and NASDAQ by Gopikrishnan et al. \cite{Gopikrishnan00}, and for the LSE and Paris by Gabaix et al. \cite{Gabaix04}.}.  Similar values are computed for individual stocks, as shown in Fig.~\ref{tailExponents}.  The average values are $\alpha = 1.74 \pm 0.23$ for off-book, $\alpha = 4.2 \pm 1.5$ for on-book, and $\alpha = 1.36 \pm 0.10$ overall.  These results are consistent with the hypothesis that order splitting is more common in the on-book market than it is in the off-book market.  However, they also suggest that the separation between the styles of trading in these two markets is not absolute.  They both show an approximate power law decay in their tails, although this decay is much steeper for the on-book market.  

Finally Fig.~\ref{tailExponents} shows that the exponent for the volume distribution of the aggregate of the on- and off- book trades is systematically smaller than the exponent for either of them by themselves. This is caused by the aggregation of two distributions:  Mixing distributions with different scaling properties tends to fatten the tails.  It indicates that one should be very careful in aggregating distributions\footnote{When power law distributions are combined the one with the lowest tail exponent determines the tail exponent of the aggregate.  For a finite sample, however, there are often slow convergence effects as a function of sample size that can alter this conclusion.}.  

\subsection{Predicted vs. actual values of $\gamma$\label{gamma}}

\begin{figure}[ptb]
\begin{center}
\includegraphics[scale=0.3,angle=-90]{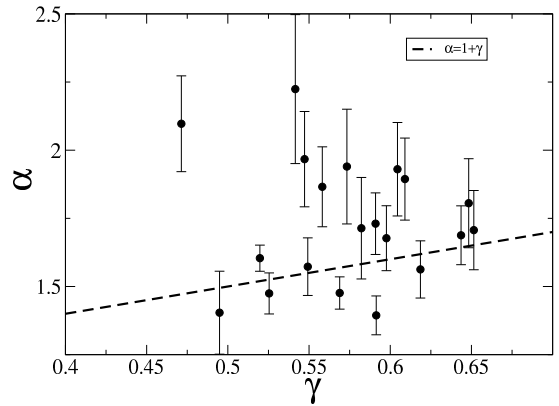}
\caption
{The scaling exponents $\alpha$ for the twenty stocks we study here (with the hypothesis $P(V > x) \sim x^{-\alpha}$), plotted against the exponent $\gamma$ of the autocorrelation function (under the hypothesis $\rho(\tau) \sim \tau^{-\gamma}$).   The error bars shown are the $95$-percent confidence intervals of the Hill estimator, under the assumptions of IID errors and perfect Pareto scaling across the entire range of $V$.  Both assumptions are highly optimistic.}
\label{alphaGamma}
\end{center}
\end{figure}

Taking the off-book market as a crude proxy for hidden orders, we test the model by comparing $\hat{\gamma} = \alpha - 1$ as predicted by Eq.~(\ref{pareto}) to the value of $\gamma$ measured directly from the order signs.  The scaling exponent $\gamma$ is measured by computing the Hurst exponent of the series of market order signs for each stock using the DFA method \cite{Peng94}, and making use of the relation $\gamma = 2(1 - H)$.   (This is much more accurate than computing the autocorrelation function directly).  We compare the predicted and actual values in Fig.~\ref{alphaGamma}.  The average value of the scaling exponent of the autocorrelation function is $\gamma = 0.57 \pm 0.05$.  This can be compared either to $\hat{\gamma} =  0.74 \pm 0.23$ based on the average value of $\alpha$, or to $\hat{\gamma} = 0.59$ based on the $\alpha$ for the aggregate distribution.  In either case the agreement is well within the error bars.  (The error bars, which are based on the standard error of the mean of the 20 stock sample, are highly optimistic due to correlations within the sample and possibly also due to skewness and systematic bias of the Hill estimates).

As a stronger test, one might hope that variations in measured values of $\alpha$ might predict variations in measured values of $\gamma$.  The model fails this test.  Performing a regression of predicted vs. actual values gives a statistically insignificant, slightly negative slope.  There are several possible explanations for this:  First, as we have already discussed, the off-book data may be a poor proxy for hidden orders.  Second, the sample errors are very large, particularly for measuring $\alpha$.  The errors bars we have shown for $\alpha$ in Fig.~\ref{alphaGamma} are the $95$-percent confidence intervals of the Hill estimator under the assumption that the data are IID and that the top one percent of the values have converged to a perfect Pareto distribution.  This is clearly far too optimistic.  This can be seen by breaking the data into subsamples; the variation from year to year is much larger than the error bars given by the Hill estimator.  Even though our samples are large, the errors are still large because both volume and order signs are long-memory processes \cite{Lobato00,Lillo03c}, and averages generally converge as $T^{-(1 - H)}$, where $H \approx 0.75$ in both cases.   In addition, the measured values of $\alpha$ have larger errors than those of $\gamma$ due to a strong tendency of the volume to trend upward, an effect that isn't easily removed by simple normalization.  Gabaix {\it et al.} have conjectured that the exponent $\alpha$ for the volume distribution has a universal value $\alpha = 3/2$; if true, this would imply that deviations from that value are purely statistical fluctuations.  Finally, it is of course possible that our model is wrong, due to violations of the assumptions of the model.  We list some of the possible problems in Section~\ref{assumptions}.

\subsection{Run length \label{runlength}}

\begin{figure}[ptb]
\begin{center}
\includegraphics[scale=0.3,angle=-90]{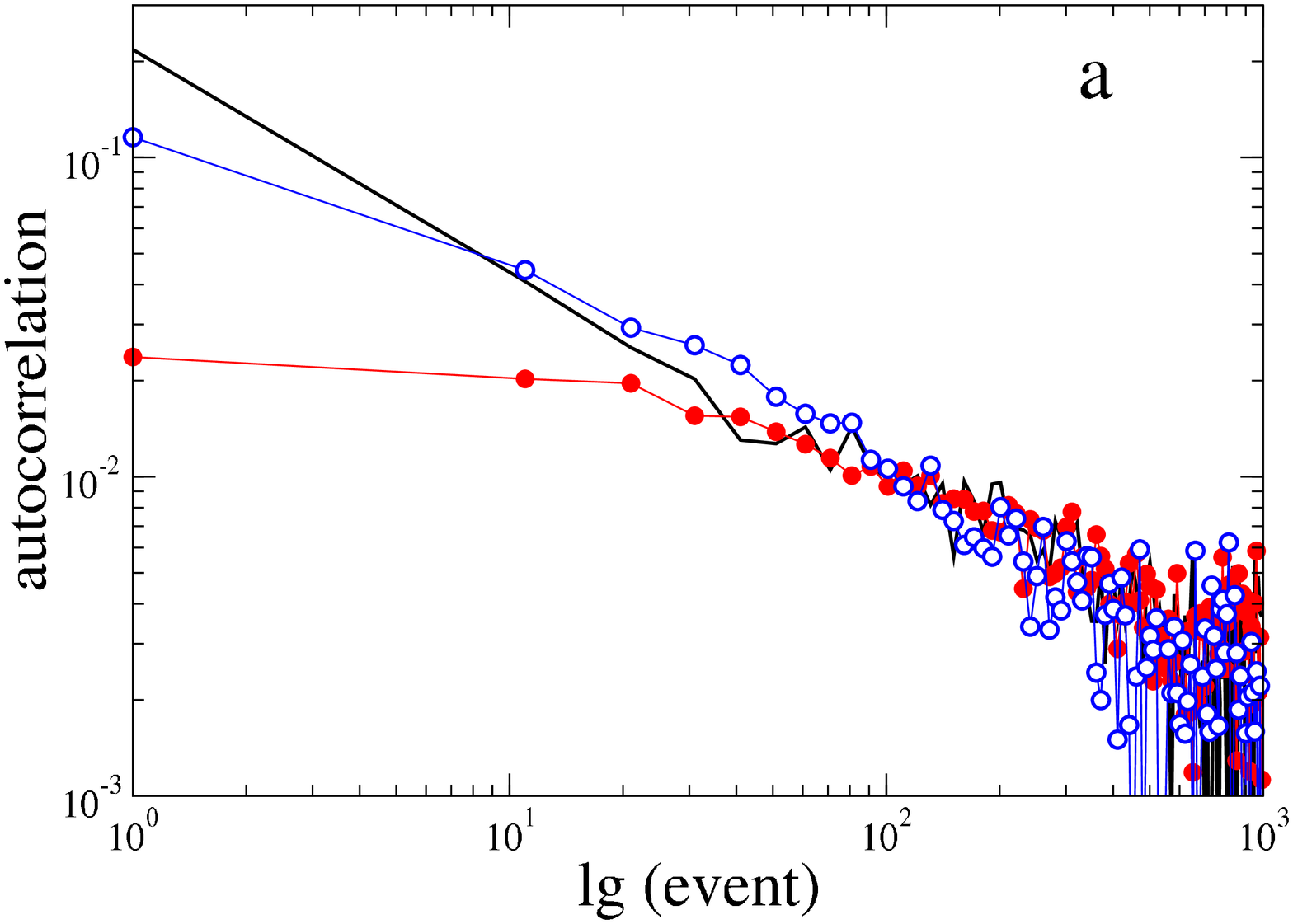}
\includegraphics[scale=0.3,angle=-90]{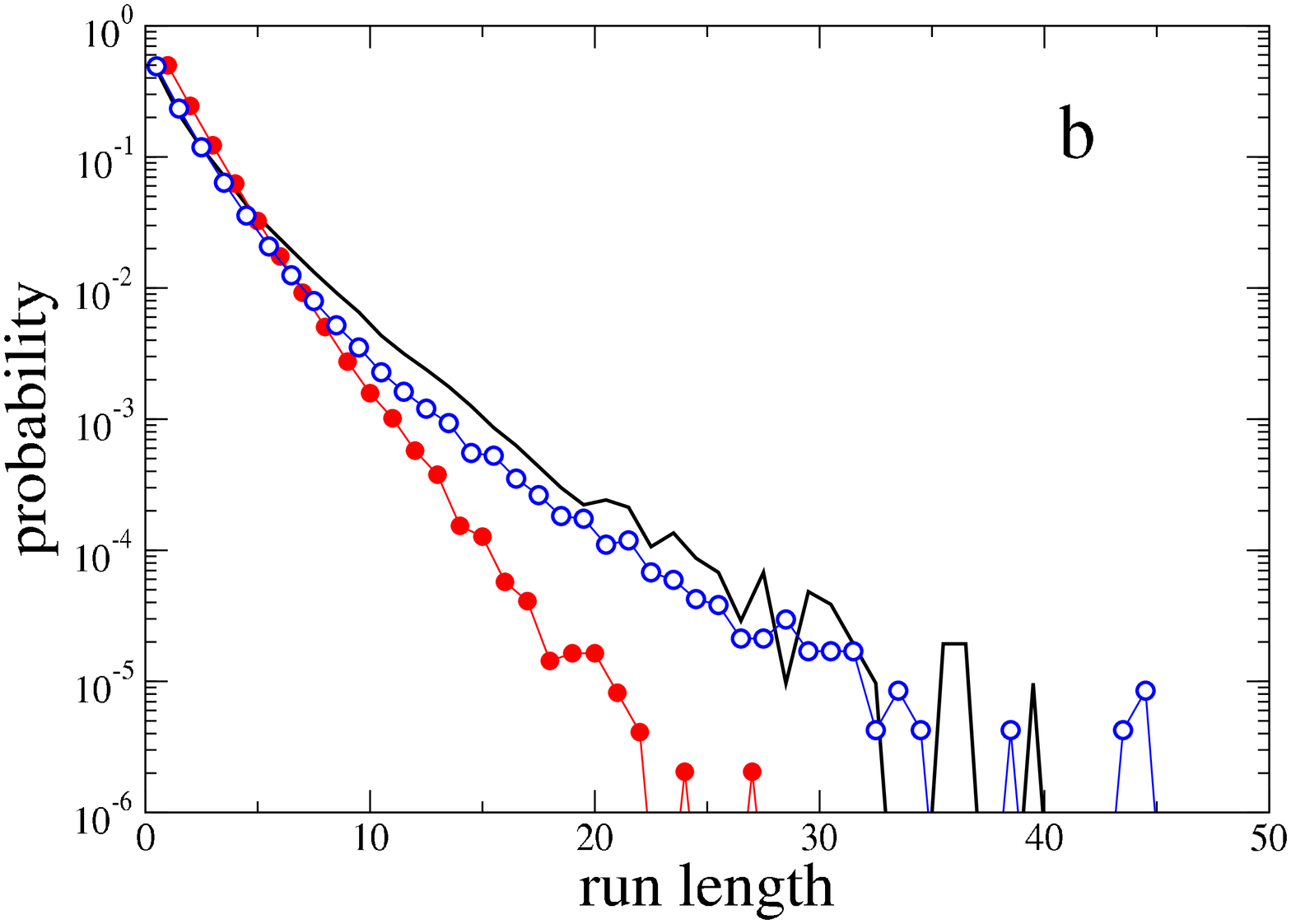}
\end{center}
\caption{(Color online) (a) Autocorrelation function of the market order sign for the stock Astrazeneca (black line) compared with the autocorrelation function of a numerical simulation of the fixed $N$ model (red filled circles, parameters $N=24$ and $\alpha=1.63$) and of the $\lambda$ model (empty blue circles, parameters $\alpha=1.63$, and $\lambda=0.38$ (which implies an average value of $N=21.1$). (b) Probability distribution of the run length for real data and simulations of the model.  The symbols and parameters are the same as in panel (a).}
\label{runs}
\end{figure}

Another test for comparing the models to data concerns the distribution of run lengths. A run is a series of revealed orders that are all of the same sign.  In figure~\ref{runs} we compare the run length distribution of the real order flow with a simulation of the both the fixed $N$ model and the $\lambda$ model. In panel (a) we show the autocorrelation function of the sign of market orders for the stock Astrazeneca (AZN) and compare it with the autocorrelation of a simulation of the two models. The parameters are $N=24$ and $\alpha=1.63$ for the fixed $N$ model and $N=21.1$, $\alpha=1.63$, and $\lambda=0.38$ for the $\lambda$ model.  These parameters were chosen to give a best fit to the autocorrelation function of the real data.  Both models are able to capture the asymptotic behavior of the autocorrelation function, but the fixed $N$ model clearly underestimates the autocorrelation function for small lags.   We can get a more detailed test by comparing the run length distribution of the models and the data, as shown in see panel (b) of figure~\ref{runs}). The figure shows that the $\lambda$ model is able to describe the run length distribution, whereas the fixed $N$ model underestimate the run length probability for long runs. The $\lambda$ model appears to be a better candidate for describing real order flow.   

\subsection{Review of assumptions \label{assumptions}}

Below we give a brief discussion of the assumptions of the model, as well as the circumstances under which this might alter the basic conclusions of the model.

\begin{itemize}
\item
{\it Distribution of hidden orders.}
This has already been discussed in some detail above. Here we want to add that we have not addressed the possible cause of the power law distribution of hidden orders.  One possibility (originally suggested by Levy and Solomon and developed by Gabaix et al.~\cite{Levy96b,Gabaix03,Levy05}) is that the hidden order size distribution is in some way related to the power law distribution of the size of holdings of the largest market participants.

\item
{\it IID hidden order arrival}.  Strong autocorrelations in hidden order size or hidden order signs could affect $\gamma$, particularly if these were strong enough to be long-memory.
\item
{\it Distribution of revealed orders}.  In reality, revealed orders do not have constant size.  If their distribution is sufficiently thin tailed we think the model should still be valid.  Power law tails, however, might affect $\gamma$.
\item
{\it Aggregation of orders.} 
In reality, there is a limited number of brokerage firms, and when they receive hidden orders with opposite signs within a sufficiently short period of time, they may cross such orders internally before they execute the remainder externally.  This will reduce the amount of unexecuted volume and improve market clearing.  In our model it has the potential to change the effective value of $N$.  However, because of the independence of the asymptotic scaling behavior on $N$, we do not think this will affect $\gamma$.
\item {\it Feedback between order execution and order generation.}  In our model we do not worry about whether revealed orders are actually executed.  In reality many revealed orders may never be executed.  In this case there may be feedback effects, i.e., if an order is not executed the hidden order size is not decreased, and consequently may result in the generation of additional revealed orders when the agent tries again.  We cannot say with certainty that such effects are not important.  However, one piece of relevant evidence is that within statistical error the same scaling is observed
for market orders, limit orders, and cancellations \cite{Lillo03c}. Since market orders are by definition executed immediately, this suggests that such feedback effects are of minor importance.

\end{itemize}

\section{Discussion}

We have presented and solved a rather idealized model of the long-memory of order flow which was designed to yield tractable results.  As detailed in the preceding section, many of its assumptions are not strictly true.  At the very least, though, it illustrates how two apparently disparate phenomena may be linked together, and makes quantitative predictions about their relationship.  Because we lack the proper data to test the model, we have used an imperfect proxy to test the model.  The model passes this test.  However, it would be nice to do a more definitive test, based on a data set that more closely characterizes the dichotomy between hidden and revealed orders.  Even if the model is not strictly true, the model could potentially be extended to include more realistic assumptions, such as a non-trivial distribution of revealed order sizes.

The long-memory of order signs is interesting for its own sake, but it may also have more profound effects on other aspects of the market.  The persistent autocorrelation function associated with a long-memory process implies a high degree of predictability by just constructing a simple linear time series model (see refs.~\cite{Bouchaud04,Lillo03c}).  Since buy orders tend to generate a positive price response, and sell orders tend to generate a negative price response, all other things being equal this would translate into easily exploitable predictable movements in prices.  In order to prevent this from happening, other features of the market have to adjust to compensate.  Such features include the size of buy vs. sell orders, the volume of  unexecuted orders at the best prices, and many other aspects of the market \cite{Bouchaud04,Lillo03c,Bouchaud04b}.  Market participants do not behave out of philanthropic motives; presumably these effects all come about due to the application of profit-making strategies.  It is not at all obvious what these strategies are, and how they combine to eliminate this inefficiency.  The market response to the long-memory of order flow is an interesting example of a self-organized collective phenomenon.  It may be one of the causes of other important properties of prices, such as the long-memory in their diffusion rate.  We have demonstrated that the $\lambda$ model, which allows fluctuations in the number of hidden orders, automatically generates fluctuations in liquidity. This is known to affect price diffusion rates \cite{Farmer04b}.  The independence on the number of hidden orders, which was not obvious to us before doing the calculation, is a convenient property of our result that makes it possible to test the model based on information that can be feasibly gathered.  This is thus a falsifiable model.

\section*{ACKNOWLEDGMENTS}
FL thanks partial funding support from research projects MIUR 449/97 ``Dinamica di altissima frequenza nei mercati finanziari" and MIUR-FIRB RBNE01CW3M.  SM and JDF would like to thank Credit Suisse First Boston, the McDonnell Foundation, Bill Miller, and Bob Maxfield for funding this work.

\section*{APPENDIX}
In this appendix we evaluate the asymptotic behavior of the autocorrelation $\rho(\tau)$ of Eq.~(\ref{final}) when the hidden order size $L$ has a Pareto distribution of Eq.~(\ref{pareto}).  We split the integral of Eq.~(\ref{final}) in three parts and we set $b=p(1-p)$.

The first contribution is
\begin{equation}
-\int_{3/2}^{\infty}\frac{p\alpha}{2\bar L L^{\alpha+1}}\sqrt{\frac{2}{\pi}}\sqrt{bs}\exp\left(-\frac{(ps-1)^2}{2bs}\right)~dL.
\end{equation}
This can be calculated explicitly.  It is  
\begin{equation}
-\frac{p}{2\bar L}\sqrt{\frac{2}{\pi}}\sqrt{bs}\exp\left(-\frac{(ps-1)^2}{2bs}\right),
\end{equation}
which asymptotically goes as
\begin{equation}
 -\sqrt{s}\exp\left(-\frac{ps}{2(1-p)}\right).
 \label{1}
\end{equation}
This decay is very fast due to the exponential term.

The second contribution is 
\begin{equation}
\frac{p\alpha}{2\bar L}\sqrt{\frac{2}{\pi}}\sqrt{bs}\int_{3/2}^{\infty}\frac{\exp\left(-\frac{(L-1-sp)^2}{2bs}\right)}{ L^{\alpha+1}}~dL.
\label{second}
\end{equation}
This integral cannot be computed analytically.  In order to get its asymptotic behavior for large $s$ (i.e. large $\tau$) we make use of the saddle point approximation \cite{Olver74}. To have an idea of the approximation let us consider the case in which one has to calculate the asymptotic behavior of an integral of the type
\begin{equation}
\int_a^b dx~e^{Nf(x)}
\label{saddle}
\end{equation}
for large values of $N$.  If there exists a point $x_0$ in $(a,b)$ which is a minimum for $f(x)$, then we can expand $f(x)$ around $x_0$, yielding
\begin{equation}
e^{Nf(x)}\simeq \exp[N(f(x_0)+\frac{1}{2}f''(x_0)(x-x_0)^2)],
\end{equation}
and we can compute the Gaussian integral 
\begin{equation}
\int_a^b dx~e^{Nf(x)}\simeq \sqrt{\frac{2\pi}{f''(x_0)}}\exp(Nf(x_0)).
\end{equation}
The method can be applied also when the integral is not of the form (\ref{saddle}), given that the integrand can be written as $\exp(f(x,N))$. In our case the integral in Eq.~(\ref{second}) can be rewritten as
\begin{equation}
\int_{3/2}^{\infty} \exp\left(-\left[\frac{(L-1-sp)^2}{2bs}+(\alpha+1)\log x\right]\right)~dL.
\end{equation}
By applying the saddle point approximation one easily gets for the integral the approximation
\begin{equation} 
\sqrt{2\pi bs} \exp\left(\frac{1}{4bs}\right) (sp)^{-(\alpha+1)},
\end{equation}
and by putting also the prefactor we get for the second contribution
\begin{equation}
(\alpha-1)p(1-p)\exp\left(\frac{1}{4bs}\right) (sp)^{-\alpha}\sim \frac{1}{\tau^{\alpha}}.
\label{2}
\end{equation}
Thus the second contribution gives a power law behavior but with an exponent $\alpha$ rather than $\alpha-1$.

The third contribution is the one depending on the three ${\rm erf}$ functions
\begin{eqnarray}
\frac{p\alpha}{2\bar L}\int_{3/2}^{\infty}(sp-1){\rm erf}(\frac{1-sp}{\sqrt{2bs}})+(L-2){\rm erf}(\frac{1/2+sp}{\sqrt{2bs}})\nonumber \\
+(1+sp-L){\rm erf}(\frac{1-L+sp}{\sqrt{2bsp}})~dL.
\end{eqnarray}
After some algebraic manipulations we can rewrite this term as
\begin{widetext}
\begin{eqnarray}
\frac{p(\alpha-1)}{2\alpha}\left(\frac{\alpha}{\alpha-1}-2\right)\left[{\rm erf}\left(\frac{1-ps}{\sqrt{2bs}}\right)+{\rm erf}\left(\frac{1/2+sp}{\sqrt{2bs}}\right)\right] \nonumber \\
+\frac{p(\alpha-1)}{2}\int_{3/2}^{\infty}(L-1-ps){\rm erf}\left(\frac{1-ps}{\sqrt{2bs}},\frac{L-1-ps}{\sqrt{2bs}}\right)\frac{1}{L^{\alpha+1}}~dL,
\label{terzo}
\end{eqnarray}
\end{widetext}
where ${\rm erf}(x_1,x_2)={\rm erf}(x_2)-{\rm erf}(x_1)$ \cite{Abramowitz74}, and we have used the fact that $\bar L = \alpha/(\alpha-1)$.
The term in square brackets has asymptotic behavior 
\begin{equation}
\frac{p(\alpha-1)}{2\alpha}\left(\frac{\alpha}{\alpha-1}-2\right)\sqrt{\frac{2bs}{\pi}}(e^{p/2b}-e^{p/b})\frac{\exp\left(-\frac{p^2s}{2b}\right)}{ps},
\label{3a}
\end{equation}
and it is dominated by the exponential. The result is obtained by using the asymptotic expansion of the ${\rm erf}$ function.

Finally we compute the asymptotic behavior of the integral in Eq.~(\ref{terzo}), i.e.
\begin{equation}
I\equiv \int_{3/2}^{\infty}(L-1-ps){\rm erf}\left(\frac{1-ps}{\sqrt{2bs}},\frac{L-1-ps}{\sqrt{2bs}}\right)\frac{1}{L^{\alpha+1}}~dL.
\end{equation}
 It is convenient to perform first an integration by parts obtaining 
 \begin{widetext}
 \begin{eqnarray}
 I=\frac{1}{L^{\alpha}}\left(\frac{L}{1-\alpha}+\frac{1+ps}{\alpha}\right){\rm erf} \left(\frac{1 - ps}{\sqrt{2bs}},\frac{L-1-ps}{\sqrt{2bs}}\right)\big|_{3/2}^{\infty}-\nonumber \\
 \int_{3/2}^{\infty}\frac{1}{L^{\alpha}}\left(\frac{L}{1-\alpha}+\frac{1+ps}{\alpha}\right) \frac{2}{\sqrt{\pi}\sqrt{2bs}}\exp\left(-\frac{(L-x-ps)^2}{2bs}\right)~dL.
 \end{eqnarray}
 \end{widetext}
The finite term decays exponentially to zero because of the properties of the error function. The asymptotic behavior of the two integrals can be computed with the saddle point method in the same way as Eq.~(\ref{second}). Both decay asymptotically as $s^{-\alpha+1}$ and the final result is
\begin{equation}
 \frac{p(\alpha-1)}{2}I\sim \frac{1}{\alpha p^{\alpha-2}}\frac{1}{s^{\alpha-1}}\sim \frac{1}{\alpha p^{\alpha-2}}\frac{1}{\tau^{\alpha-1}},
 \end{equation}
which coincides with Eq.~(\ref{paretoScaling}).

\bibliographystyle{plainnat}
\bibliography{jdf} 

\end{document}